\documentclass[iop]{emulateapj}
\usepackage{epsfig}
\usepackage{apjfonts}
\usepackage{aas_macros}
\usepackage{amsmath}
\usepackage{graphicx}

\begin{document}

\title{A Possible kilonova powered by magnetic wind from a newborn black hole}

\author{Shuai-Bing Ma\altaffilmark{1}, Wei Xie\altaffilmark{2},  Bin Liao\altaffilmark{1}, Bin-Bin Zhang\altaffilmark{3,4}, Hou-Jun L\"{u}\altaffilmark{5}, Yu Liu\altaffilmark{1}and Wei-Hua Lei*\altaffilmark{1}}

\altaffiltext{1}{Department of Astronomy, School of Physics, Huazhong University of Science and Technology,  Wuhan, 430074, China}
\altaffiltext{2}{Guizhou Provincial Key Laboratory of Radio Astronomy and Data Processing,  Guizhou Normal University, Guiyang, 550001, People's Republic of China}
\altaffiltext{3}{School of Astronomy and Space Science, Nanjing University, Nanjing 210093, China}
\altaffiltext{4}{Key Laboratory of Modern Astronomy and Astrophysics (Nanjing University), Ministry of Education,  China}
\altaffiltext{5}{Guangxi Key Laboratory for Relativistic Astrophysics, School of Physical Science and Technology, Guangxi University, Nanning 530004, China}
\email{leiwh@hust.edu.cn}

\begin{abstract}

The merger of binary neutron stars (NS-NS) as the progenitor of short Gamma-ray bursts (GRBs) has been confirmed by the discovery of the association of the gravitational wave (GW) event GW170817 with GRB 170817A. However, the merger product of binary NS remains an open question. An X-ray plateau followed by a steep decay (``internal plateau'') has been found in some short GRBs, implying that a supra-massive magnetar operates as the merger remnant and then collapses into a newborn black hole (BH) at the end of the plateau. X-ray bump or second-plateau following the ``internal plateau'' is considered as the expected signature from the fallback accretion onto this newborn BH through Blandford-Znajek mechanism (BZ). At the same time, a nearly isotropic wind driven by Blandford-Paynemechanism (BP) from the newborn BH's disk can produce a bright kilonova. Therefore, the bright kilonova observation for a short GRB with ``internal plateau'' (and followed by X-ray bump or second-plateau) provides further evidence for this scenario. In this paper, we find that GRB 160821B is a candidate of such a case, and the kilonova emission of GRB 160821B is possibly powered by the BP wind from a newborn BH. Future GW detection of GRB 160821B-like events may provide further support to this scenario, enable us to investigate the properties of the magnetar and the newborn BH, and constrain the equation of state of neutron stars.

\end{abstract}

\keywords{accretion, accretion disks - black hole physics - gamma-ray burst: individual (GRB 160821B)}

\section{Introduction}
Gravitational wave (GW) event GW170817, detected by the LIGO-Virgo collaboration on 17 August 2017,  is generally believed from the merger of binary neutron stars (NSs)\citep{Abbott+2017a}.  The expected electromagnetic (EM) counterparts of this GW event were well confirmed by the performance of many follow-up observations in various wavelengths \citep{Abbott+2017b}. A faint short gamma-ray burst (sGRB) 170817A \citep{Zhang+2018Nat}, observed by the Fermi and INTEGRAL satellites just 1.7s after the GW detection, revealed that at least some short GRBs might be originated from NS-NS merger \citep{Goldstein+2017,Savchenko+2017}.
After $\sim 11$ hr of the merger, an ultraviolet-optical-infrared (UVOIR) transient was first found by the Swope Supernovae Survey (or 2017gfo) \citep{Arcavi+2017,Coulter+2017,Drout+2017,Kasen+2017,
Pian+2017,Smartt+2017,Soares+2017,Valenti+2017}. In X-ray and radio bands, the EM counterparts were subsequently detected \citep{Troja+2017,Dobie+2018}. This GW event associated with EM counterparts marked the beginning of the era for multi-messenger astronomy.

There are several evolutionary paths of the post-merger of GW170817-like events (Rosswog et al. 2000; Lasky et al. 2014; Gao et al.2016): (1) collapse into a black hole (BH) immediately; (2) a short-lived hyper-massive NS supported by differential rotation, which collapses into a BH within a timescale of seconds due to the combination of magnetic breaking and viscosity; (3) a temporal supra-massive millisecond magnetar supported by rigid rotation, which collapses to a BH at a later time after spinning down; and (4) a stable NS. However, limited to the sensitivity of gravitational wave detectors, the product (a NS or a black hole) of the NS-NS merger is poorly constrained.

From the observational point of view, the afterglows of short GRBs may leave some clues for the merger product. \citet{Troja+2007} first pointed out that the X-ray ``internal plateau'' followed by a sharp decay are likely powered by a magnetar. The sharp decay at the end of the plateau, marking the abrupt cessation of the magnetar's central engine, has then been naturally interpreted as the collapse of a supra-massive magnetar into a black hole \citep{Troja+2007,Rowlinson+2013,Lv+2015}. If this ``internal plateau'' is indeed an evidence of a magnetar, the signatures from the new-born BH would at least be expected to leave in some GRBs. The small X-ray bump following the ``internal plateau'', first uncovered by \citet{Troja+2007}, can be well explained with a fallback accretion \citep{Chen+2017}. Another example, but for short GRB, is GRB 110731A. Recently, \citep{Zhao+2020} found that some GRBs (GRBs 070802, 090111, and 120213A) show a second plateau in X-ray afterglow. These X-ray bumps or second X-ray plateaus are considered as a possible signature of the newborn BH \citep{Chen+2017,Zhang+2018,Zhao+2020}.

On the other hand, the ``kilonova'' powered by radioactive decay of the r-process \citep{Lilx+1998,Metzger+2010} has been broadly adopted to interpret AT 2017agfo, the optical counterpart of GW170817 \citep{Kasen+2017}.
If the supra-massive NS (as the product of NS-NS merger) finally collapses into a newborn BH, the magnetic wind driven by Blandford-Payne process (BP) from the BH accretion disk would heat up the neutron-rich merger ejecta and produce a more bright kilonova. Therefore, besides the X-ray bump or second X-ray plateau, a bright kilonova powered by BP driven wind is also expected in this scenario, and thus provide a new signature of the new-born BH.

In this paper, we find that a nearby short GRB 160821B is a possible candidate showing these two kinds of features. The ``internal plateau'' found in X-ray afterglow implies a supra-massive magnetar operating after the merger of double NSs \citep{Lv+2017, Kisaka+2017}. The second X-ray plateau following the ``internal plateau'' in GRB 160821B can be well interpreted with a fallback BH, indicating that a newborn BH is likely active \citep{Kisaka+2017,Zhang+2018}. The optical/nIR observations show a clear evolution toward red colors, which has been considered as a well-sampled kilonova \citep{Troja+2019a, Lamb+2019, Yuan+2020}.
We find that the kilonova in GRB 160821B is too bright to be powered by r-process alone, but is more consistent with a kilonova powered by the magnetic wind from the newborn BH. These results suggest that a newborn BH from the collapse of a supra-massive magnetar may indeed come into play, leading to a candidate kilonova powered by the magnetic wind from this newborn BH. Therefore, the observations of GRB afterglow as well as kilonova may contain clues of the merger product of NS-NS binaries.

The paper is organized as follows: In Section 2, we describe the spin-down of a supra-massive magnetar and the fallback accretion of BH central engine. We also study the effects of energy injection from BH on GRB afterglow and kilonova, respectively. In Section 3, we apply our model to explain the multi-band afterglow of GRB 160821B and its accompanied kilonova. Finally, we briefly summarize our results and discuss the implications in Section 4.

\section{The Model}
The associations of GW170817 with GRB 170817A indicates that a fraction of short GRBs might have a NS-NS merger origin. If the merger product of NS-NS binaries is a supra-massive magnetar supported by centrifugal force, it may collapse to a newborn BH when it spins down. The ``internal plateaus'' observed in some short GRBs are believed to be powered by a magnetar central engine. The sharp decay followed the ``internal plateau'' marks the birth of a BH from the collapse of magnetar.

The fallback accretion onto the newborn BH would leave some imprints on GRB afterglow and kilonova. The jet power produced by Blandford-Znajek mechanism (BZ) will be continuously injected into the external shock during the fallback accretion stage, resulting in flattening in the light curves of afterglow. As discussed in Ma et al. (2018), a near-isotropic wind driven by the BP mechanism can heat up and push the merger ejecta, leading to a more bright kilonova. The sketch of BZ powered jet and BP driven kilonova are illustrated in Figure 1.

\begin{figure}[htp]
\center
\includegraphics[width=0.4\textwidth]{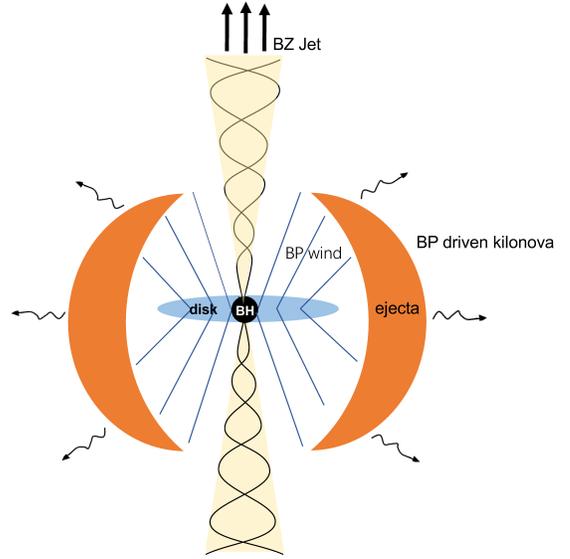}
\caption{Illustration of the BZ and BP mechanisms. The energy extraction from the BH via BZ mechanism can affect GRB afterglow. Simultaneously, a magnetic wind is driven from the fallback accretion disk by the BP mechanism. A fraction of the BP energy is dissipated into the merger ejecta, which would produce a bright kilonova around days. The coexistence of a BZ jet and a BP wind may develop a structured jet as expected in GRB 170817A \citep{Mooley+2018,Troja+2019b}.}
\end{figure}

\subsection{Spin down of a supra-massive magnetar}
The supra-massive millisecond magnetar, as the product of NS-NS merger, would spin down due to its rotation energy loss through both magnetic dipole radiation and gravitational wave (GW) radiation~\citep{Zhang+2001}. The spin-down rate can be expressed as
\begin{equation}
\dot{E}=I\Omega \dot{\Omega}=-\frac{B_p^2R^6\Omega^4}{6c^3}-\frac{32GI^2 \epsilon ^2 \Omega^6}{5c^5},
\label{eq:Edot}
\end{equation}
where $I$ is the moment of inertia, $\Omega=2\pi/P$ is the angular frequency and $\dot{\Omega}$ is the time derivative, $B_{\rm p}$ is the surface magnetic field strength at the poles, $R$ is the radius of the magnetar and $\epsilon$ is the ellipticity of the magnetar.

Considering that the spin-down of magnetars through magnetic dipole radiation and GW radiation,
by using Eq.(\ref{eq:Edot}) one can define two characteristic spin-down time scales
\begin{eqnarray}
t_{\rm md} & = & \frac{3 c^3 I}{B_p^2 R^6 \Omega_0^2} \simeq 2.0 \times 10^3~{\rm s}~(I_{45} B_{p,15}^{-2} P_{0,-3}^2 R_6^{-6}), \label{eq:tmd}\\
t_{\rm GW} & = & \frac{5c^5}{128 GI \epsilon^2 \Omega_0^4} \simeq 9.1\times 10^3~{\rm s}~(I_{45}^{-1} P_{0,-3}^4 \epsilon_{-3}^{-2}),
\end{eqnarray}
where $\Omega_0$ and $P_0$ are the initial angular velocity and period of the magnetar, respectively.
The overall spin-down time scale of the magnetar
can be defined as
\begin{equation}
t_{\rm sd} = {\rm min} (t_{\rm md}, t_{\rm GW}),
\label{eq:tsd}
\end{equation}

Since the ellipticity of magnetar is unknown, we consider only the energy loss due to dipole radiation in this work. As a result, the supra-massive magnetar can produce an ``internal plateau'' in the X-ray afterglow of GRB. The characteristic spin-down luminosity due to dipole radiation is $L_{\rm md} = 1.0\times 10^{49}~{\rm erg~s}^{-1}(B_{\rm p,15}^2 P_{0,-3}^{-4}R_6^6) $.
The isotropically equivalent luminosity of the ``internal plateau'' ($L_{\rm X}$) is interpreted with the spin-down luminosity  $L_{\rm md}$ as
\begin{equation}
L_{\rm X}\simeq L_{\rm md}. \label{Lint}
\end{equation}

The evolution of magnetar spin period due to dipole radiation is given by
\begin{equation}
P(t)=P_0\left(1+\frac{t}{t_{\rm md}}\right)^{1/2}. \label{Pt}
\end{equation}

The supra-massive magnetar collapses to a BH when its spin period $P$ becomes large enough that $M_{\rm max}(P) = M_{\rm p}$, resulting a sharp drop at the end of ``internal plateau''. $M_{\rm p}$ is the mass of the protomagnetar. The maximum gravitational mass ($M_{\rm max}$) of a supra-massive magnetar can be writen as a function of the spin period \citep{Lasky+2014},
\begin{equation}
M_{\rm max}=M_{\rm TOV}\left(1+\hat{\alpha} P^{\hat\beta}\right),
\label{Mmax}
\end{equation}
where $M_{\rm TOV}$ is the maximum mass for a non-rotating NS, $\hat\alpha$ and $\hat\beta$ depend on the equation of state (EOS).  Recent studies with short GRB data favor the EoS GM1 ($M_{\rm TOV}=2.37~M_{\sun}$, $R=12.05~{\rm km}$,
$I=3.33\times10^{45}~{\rm g~cm}^{-2}$, $\hat\alpha=1.58\times10^{-10}{\rm s}^{-\hat\beta}$ and $\hat\beta=-2.84$)  \citep{Lv+2015, Lv+2017}. In this paper, we adopt the EOS GM1.

The collapse time $t_{\rm col}$ of a supra-massive magnetar is a function of the EOS \citep{Sarin+2020}. As discussed in \citet{Lv+2015}, for the ``internal plateau'' with  post-break decay slope $\alpha \geq 3$, $t_{\rm col}$ is just defined by the observed
break time $t_{\rm b}$ as
\begin{equation}
t_{\rm col}= t_{\rm b} / (1+z) \leq t_{\rm md} \label{tcol}
\end{equation}
In such case, we can derive the upper limit for $P_0$ and $B_{\rm p}$ by combing the equations \eqref{Lint} and \eqref{tcol}.

Assuming that the newborn BH inherits the mass and angular momentum from
the supra-massive magnetar, one can estimate the initial BH mass as $M_\bullet= M_{\rm p}$ (the mass of the protomagnetar $M_{\rm p}$ equals the critical mass $M_{\rm max}$ at $t_{\rm col}$) and initial angular moment as $J_\bullet= 2\pi I/P_0$.

\subsection{Fallback accretion onto the newborn BH}
As the collapse of magnetar, some of the surrounding matter initially blocked by the magnetic barrier would begin to fallback and be accreted by the newborn BH. A relativistic jet can be then launched via the BZ mechanism \citep{Blandford+1977}, in which the spin energy of the newborn BH is extracted through the open field lines penetrating the event horizon. For a BH with mass $M_\bullet$ and spin $a_\bullet$ ($\equiv J_\bullet c/(GM_\bullet^2)$ ), the BZ power can be estimated as \citep{Lei+2013,Lei+2017,Liu+2017},
\begin{equation}
L_{\rm BZ}=1.7 \times 10^{50}  a_{\bullet}^2 m_{\bullet}^2
B_{\bullet,15}^2 F(a_{\bullet}) \ {\rm erg \ s^{-1}},
\end{equation}
where $m_\bullet = M_\bullet/M_\sun$, $F(a_\bullet) = [(1+q^2)/q^2] [(q+1/q)\arctan q-1]$, and $q=a_\bullet/(1+\sqrt{1-a_\bullet^2})$. $B_{\bullet,15}=B_{\bullet}/10^{15} {\rm G}$, and $B_\bullet$ is the magnetic field strength threading the BH horizon.

As the magnetic field on the BH is supported by the surrounding disk, it is reasonable to assume that the magnetic pressure on the horizon may reach a fraction $\alpha_{\rm m}$ of the ram pressure of the innermost parts of an accretion flow, i.e.,
\begin{equation}
B_\bullet^2 /8\pi = \alpha_{\rm m} P_\mathrm{ram} \sim \alpha_{\rm m} \dot{M} c/(4\pi r_\bullet^2).
\label{B2}
\end{equation}
where $\dot{M}$ is the disk accretion rate onto BH. We assume that the disk accretion at late time just tracks the fallback accretion rate \citep{Dai+2012,Wu+2013},
\begin{eqnarray}
\dot{\ M} = \dot{ M}_{\rm p} \left[ \frac{1}{2}\left(\frac{t-t_0}{t_{\rm p}-t_0} \right)^{-1/2} +  \frac{1}{2}\left(\frac{t-t_0}{t_{\rm p}-t_0} \right)^{5/3} \right]^{-1},
\label{dotm14}
\end{eqnarray}
where $t_0$ is the beginning time of the fallback accretion, $t_{\rm p}$ is the time corresponding to the peak fallback rate $\dot{ M}_{\rm p}$.

At the same time, A baryon-rich wide wind/outflow is then centrifugally launched by the BP mechanism \citep{Blandford+1982}. In this mechanism, energy and angular momentum are extracted magnetically from the accretion disks by the large-scale field lines threading the disk. The magnetic wind power can be estimated by \citep{Livio+1999,Meier+2001}
\begin{equation}
L_\mathrm{BP}=(B_\mathrm{ms}^\mathrm{p})^2r_\mathrm{ms}^4\Omega_\mathrm{ms}^2/32c
\label{eq:Lbp}
\end{equation}
\noindent where $\Omega_\mathrm{ms}$ is the Keplerian angular velocity at the marginally stable orbit $r_\mathrm{ms}$. The expression for $r_{\rm ms}$ is given by \citep{Bardeen+1972}
\begin{eqnarray}
r_{\rm ms}/r_{\rm g} =  3+Z_2 -\left[(3-Z_1)(3+Z_1+2Z_2)\right]^{1/2},
\label{rms}
\end{eqnarray}
for $0\leq a_{\bullet} \leq 1$, where $Z_1 \equiv 1+(1-a_{\bullet}^2)^{1/3} [(1+a_{\bullet})^{1/3}+(1-a_{\bullet})^{1/3}]$, $Z_2\equiv (3a_{\bullet}^2+Z_1^2)^{1/2}$. The Keplerian angular is given by
\begin{equation}
\Omega_\mathrm{ms}=\left(\frac{GM_\bullet}{c^3}\right)^{-1}\frac{1}{\chi^{3}_\mathrm{ms}+a_\bullet},
\label{omegams}
\end{equation}
where $\chi_\mathrm{ms}$ is defined as $\chi_\mathrm{ms}\equiv\sqrt{r_\mathrm{ms}/r_\mathrm{g}}$, and $r_\mathrm{g}\equiv G M_\bullet/c^2$.

Following Blandford \& Payne (1982), the disk poloidal magnetic field $B_\mathrm{ms}^\mathrm{p}$ at $r_\mathrm{ms}$ can be expressed as

\begin{equation}
B_\mathrm{ms}^{\rm{p}} = B_\bullet (r_\mathrm{ms}/r_\bullet)^{-5/4}
\label{Bms}
\end{equation}
\noindent where $r_\bullet = r_{\rm g} (1+\sqrt{1-a_\bullet^2})$ is BH horizon radius.

\subsection{External shock with energy injection}
Due to energy injection from the BZ jet during the fallback accretion onto the newborn BH, the blast wave energy continuously increases with time, which may produce a second plateau following the steep decay.

We investigate a jet with isotropic energy $E_{\rm k,iso}$ and opening angle $\theta$ propagating into a medium with proton number density $n$. The dynamical evolutions of jet are described in \citep{Huang+2020}. Here, we consider continuously energy injected into the external shock after the birth of BH. The total kinetic energy of jet could be expressed as $E_{\rm tot}= E_{\rm 0}+ E_{\rm inj}$, where $\ E_{\rm inj}= \int L_{BZ} dt$ is the injected energy by BZ mechanism. The dynamical equation of the bulk Lorentz factor $\Gamma$ can be then written as

\begin{equation}
\frac{d\Gamma}{dm_{sw}}=-\frac{1}{M_{0}+2\Gamma m_{\rm sw}}[\Gamma^{2}-1-\frac{L_{\rm BZ}}{c^{2}} \frac{dt}{dm_{\rm sw}}]
\end{equation}
where $m_{\rm sw}$ is the swept-up mass by shock, $M_{0}$ is the initial mass of the jet.

The forward shock may be continuously refreshed with the BZ power $L_{\rm BZ}$. The electrons are believed to be accelerated
at the forward shock front to a power-law distribution $N(\gamma_{\rm e}) \propto \gamma_{\rm e}^{-p}$. A fraction $\epsilon_{\rm e}$ of the shock energy is distributed into electrons, while another fraction $\epsilon_{\rm B}$ is in the magnetic field generated behind the shock. Therefore, the emission from the electrons accelerated by the refreshed shock can be shallower or a plateau.

\subsection{Kilonova with energy injection}
Numerical simulations show that the merger of neutron star binaries (NS-NS binaries and NS-BH binaries) would lead to the formation of an ejecta with mass up to $M_{\rm ej} \sim 0.1 M_\sun$ \citep{Hotokezaka+2013,Sekiguchi+2015,Sekiguchi+2016,Dietrich+2017,Shibata+2017,Shibata+2019}. A kilonova would be expected due to the radioactive decay of the r-process from this neutron-rich ejecta.

In the case of a newborn BH with fallback accretion, the BP driven wind can heat up and push the merger ejecta as an additional energy source \citep{Ma+2018}.

The dynamic evolution of the ejecta is then given by
\begin{equation}
\frac{d\Gamma}{dt}=\frac{ L_{\rm inj}-L_{e}-\Gamma{\cal D}(dE'_{\rm int}/dt')}{M_{\rm ej}c^2+E'_{\rm int}}.
\end{equation}
where $ {\cal D}=1/[\Gamma(1-\beta)] $ is the Doppler factor, and $ \beta=\sqrt{1-\Gamma^{-2}}$.  $E_{\rm int}^\prime$ is the internal energy measured in the comoving rest frame. $L_{\rm inj}=\xi L'_{\rm BP}+L'_{\rm ra}$ denotes the injected power from a BP wind $\xi L'_{\rm BP}$ and from an radioactive decay rate $L'_{\rm ra}$. Here we introduce a parameter $\xi$ to describe the fraction of magnetic wind energy that is used to heat the ejecta. The case with $\xi=0$ or $\alpha_{\rm m}=0$ will return to the result of kilonova.

The change of the internal energy of the ejecta is thus written as
\begin{equation}
\frac{dE'_{\rm int}}{dt'}=L_{\rm inj}  -L'_e-P'\frac{dV'}{dt'}
\end{equation}
The co-moving luminosities are defined as $ L'=L/{\cal D}^2 $, and
\begin{eqnarray}
\lefteqn{L'_{\rm ra}=4\times 10^{49}M_{\rm ej,-2}}
\nonumber\\
& & \times\left[\frac{1}{2}-\frac{1}{\pi}\arctan\left(\frac{t'-t'_0}{t'_{\sigma}} \right) \right]^{1.3}{\rm erg~s^{-1}}.
\end{eqnarray}
with $t'_{0}\sim 1.3$ $\rm s$ and $t'_{\sigma}\sim 0.11$ \citep{Korobkin+2012}. For a relativistic gas, the pressure is (1/3) of the internal energy density, i.e., $P'=E'_{\rm int}/(3V')$.

The co-moving frame bolometric emission luminosity of the heated electrons can be estimated as
\begin{equation}
L'_e=\left \{ \begin{array}{ll}
E'_{\rm int}c/(\tau R/\Gamma), & \textrm{ $t < t_{\tau}$},\\
\\
E'_{\rm int}c/(R/\Gamma), & \textrm{ $t \geq t_{\tau}$},\\
\end{array} \right.
\end{equation}
where $\tau=\kappa (M_{\rm ej}/V')(R/\Gamma)$ is optical depth and $\kappa$ is opacity of the ejecta. $t_{\tau}$ is the time when $\tau=1$.

Assuming a blackbody spectrum for the thermal emission of the merger-nova, the observed flux for a given frequency $\nu$ could be calculated as
\begin{eqnarray}
F_{\nu}={1\over4\pi D_L^2}{8\pi^2  {\cal D}^2R^2\over
h^3c^2\nu}{(h\nu/{\cal D})^4\over \exp(h\nu/{\cal D}kT'_{\rm eff})-1},
\end{eqnarray}
where $D_L$ is the luminosity distance, $T'_{\rm eff}=(E'_{\rm int}/aV'\max(\tau,1))^{1/4}$ is the effective temperature, $h$ is the Planck constant, $k$ is the Boltzmann constant and $a$ is the radiation constant.

\begin{figure}[htp]
\center
\includegraphics[width=0.45\textwidth]{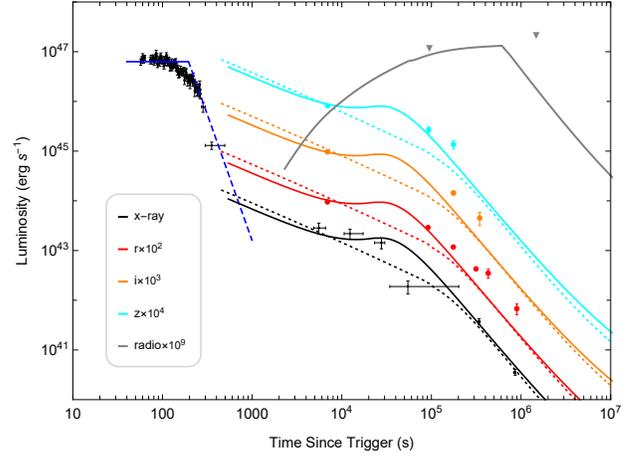}
\caption{Modeling results for the multi-band light curve of GRB 160821B. The X-ray data are detected by \textit{Swift}/XRT and XMM-Newton satellites (black). The Optical/nIR data (red,orange,cyan) and radio upper limits (gray) come from \citep{Troja+2019a}. The blue lines  show the interpretation of ``internal plateau'' with the spin-down of supra-massive magnetar.
The solid lines in different colors represent the external shock emissions with continuously energy injection through BZ process.
For comparison, we also show the results from a standard afterglow model (dotted lines) without energy injection.
The adopted parameters are: $P_0=55.8$~ms, $B_{\rm p}=1.81\times10^{17}$~G,  $a_\bullet=0.01$,
$\dot{ M}_{\rm p}=8\times10^{-10}M_{\sun}$.
\label{160821B}}
\end{figure}

\section{GRB 160821B}
GRB 160821B, with duration $T_{90} \simeq 0.5$ s and redshift $\rm z \simeq 0.1613 $ \citep{Troja+2019a}, was triggered  by the \textit{Swift} BAT at 22:29:13 UT on 2016 August 21 \citep{Siegel+2016}. Based on the duration, it was classified as a short GRB. Its short duration and low redshift made it a good candidate for kilonova searches. The optical afterglow of GRB 160821B was first detected by Swift/Ultraviolet and Optical Telescope (UVOT) in the white u filters, and subsequently detected by several ground-based optical telescopes. \citet{Troja+2019a} has collected all the optical/nIR data. In Figure~\ref{160821B}, we present the X-ray and optical (r, i and z band) data.

The X-ray afterglow of GRB 160821B showed an ``internal plateau''($L_{\rm X}\sim 10^{47}~\rm erg/s$) followed by a steep decay. The plateau extends to about $10^2$ s before rapidly falling off with a decay index $\alpha \sim 5$ (see the blue dashed line in Figure~\ref{160821B}), which is interpreted as the abrupt cessation of the central engine. The observed break time is $t_{\rm b}\sim 250$s. At the end of the sharp decay, the X-ray light curve shows a second plateau ($\alpha \sim 0.4$) followed by normal decay with power-law index $\alpha \sim 1.36$ \citep{Lv+2017,Zhang+2020}.

Considering $L_{\rm md}\simeq L_{\rm X} \simeq 10^{47}~\rm erg/s$, $t_{\rm md} \geq t_{\rm b}/(1+z)\simeq 216$s and adopting EoS GM1, we can get the upper limit for $P_0 = 55.8$~ms and $B_{\rm p}=1.81\times10^{17}$~G.
By using equation~(\ref{Mmax}), we get the lower limit of the mass of the supra-massive magnetar $M_{\rm p} = M_{\rm max}(P) \simeq 2.37 M_\sun$. The second plateau requires an extra energy input \citep{Zhang+2018,Zhao+2020}. Since the magnetar has already collapsed into a BH at this stage, a highly possible energy source would be the fallback accreting newborn BH with initial BH mass $M_\bullet \simeq 2.37 ~M_\sun $. Considering the BH angular momentum $J_\bullet= 2\pi I/P_0$, we can get the lower limit of the initial spin $a_\bullet \simeq 0.01$.

We use a numerical code developed for the external shock, and modify it to incorporate energy injection from BZ process (Wang et al. 2014). As shown in  Figure~\ref{160821B} (solid lines) , the X-ray data at $t>1000$s (i.e., the second plateau and followed normal decay) as well as the first epoch of optical observations ( at $t\sim 10^4 $s, the contribution from kilonova could be ignored at this epoch) can be well interpreted with our external shock model with energy injection from the newborn BH. The radio upper limits are used to narrow down the parameter space. The parameters we adopted are exhibited in Table 1. The BH central engine starts at $t_{0}\simeq 10^3$ s, peaks at $t_{\rm p} \simeq 8\times 10^4$ s,  and ceases at $t_{\rm f} \simeq 10^5$ s. The peak accretion rate is $\dot{M}_{\rm p} \sim 8 \times 10^{-10} ~M_\sun$. The energy from the newborn BH is injected to a jet with initial isotropic energy $E_{\rm k,iso}=1.6\times10^{50}\rm erg $ and opening angle  $\theta_{\rm j}$=3.2 deg. The jet sweeps a medium with constant number density $n_{0}=1\times10^{-4}\rm cm^{-3}$ . The electrons in the shocked region are accelerated to a power-law distribution with electron spectrum index $p=2.1$. The energy fractions of electrons and magnetic field are $\epsilon_{\rm e}$=0.2 and $\epsilon_{\rm B}$=0.1, respectively. For comparison, we also plot the the standard afterglow model (dotted lines) without energy injection, which can not account for the second plateau feature observed in X-ray afterglow.
The parameters we used are $E_{\rm k,iso}=2\times10^{50}\rm erg $, $\theta_{\rm j}$=6.0 deg, $n_{0}=1\times10^{-4}\rm cm^{-3}$, $\epsilon_{\rm e}$=0.2, $\epsilon_{\rm B}$=0.1.

\begin{table}
\begin{center}{\scriptsize
\caption{Parameters adopted for interpreting the broadband data of GRB\,160821B.}
\label{tab:parametes}
\begin{tabular}{cccccc}\hline\hline

 \multicolumn{6}{c}{Magnetar and BH Parameters}\\
  \hline
  $M_{\rm NS}~({M_\sun})$ & $B_{\rm p, 17}$ & $P_{0, -3}$ & $M_\bullet (M_{\odot})$ & $a_\bullet$ & $\dot{M}_{\rm p}~({ M_{\odot}}\rm s^{-1})$  \\
   2.37 &$1.81 \downarrow$ & $55.8 \downarrow$ & $2.37$  & $0.01\uparrow$  &  $8.0\times10^{-10}$    \\
  \hline

  \multicolumn{6}{c}{GRB afterglow parameters}\\
  \hline
  $E_{\rm k, iso}~(\rm erg)$  &  $n_{0}~({\rm cm^{-3}})$ & $\theta~({\rm deg})$ & $\epsilon_{\rm e}$ & $\epsilon_{\rm B}$ &  $p$  \\
   $1.6\times10^{50} $     &  $1\times 10^{-4}$   & $3.2 $ &  $0.2$     &  $0.1$ &  $2.1$  \\
  \hline

  \multicolumn{6}{c}{Kilonova parameters}\\
  \hline
  &$M_{\rm ej}~({\rm M_{\odot}})$  & $\kappa~(\rm cm^{2}~\rm g^{-1})$ & $ \beta~(v/c)$    & $\xi$    &   \\

  &0.011 &   $4.0$ &   $0.11 $   & $ 0.3$ &   \\
  \hline\hline

    \multicolumn{6}{c}{Other parameters}\\
  \hline
  &   $\alpha_{\rm m}$  & $t_{\rm 0}~({\rm s})$    & $t_{\rm p}~({\rm s})$ & $t_{\rm f}~({\rm s})$&  \\

  &   $0.1$   & $1\times 10^{3}$   &   $8\times 10^{4}$  & $1\times 10^{5}$ &  \\
  \hline\hline

 \end{tabular}
 }
\end{center}
\end{table}

\begin{figure}[t]
\begin{center}
\begin{tabular}{ll}
\resizebox{75mm}{!}{\includegraphics[]{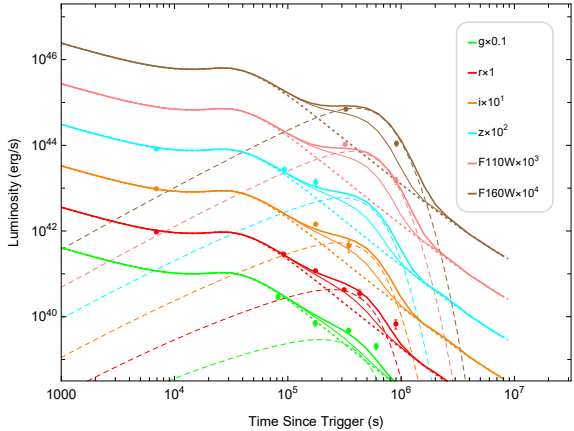}}
\end{tabular}
\caption{Multi-color optical light curves of GRB 160821B. The data show clear excess relative to the expected emissions from the GRB afterglow (dotted lines). These excess are considered as the contribution of kilonova. To interpret the data, we introduce the BP driven kilonova model (shown as dashed lines). The combined contributions from both afterglow and BP driven kilonova are plotted with thick solid lines, which are in good agreement with the data. The thin solid lines represent the emissions from the afterglow plus a r-process powered kilonova, which are disfavored by the data.}
\label{fig:KN}
\end{center}
\end{figure}

Compared to the predicted afterglow emissions, the observed optical data show clear excesses at $t> 10^4$s,  which are believed to be the contribution of kilonova \citep{Kasliwal+2017, Jin+2018, Troja+2019a, Lamb+2019, Acciari+2020, Yuan+2020}. The combined emissions (thick solid lines) from afterglow (dotted lines) and kilonova (dashed lines) are plotted in Figure~\ref{fig:KN}. As discussed in our previous work~\citep{Ma+2018}, the magnetic wind from the disk of the newborn BH with fallback accretion would heat up the neutron-rich merger ejecta and produce a bright kilonova. To see whether this effect operates in the kilonova of GRB 160821B, we represent the resulting light curves with a r-process powered kilonova and a BP-powered kilonova, denoted by thin solid lines and thick solid lines respectively. The adopted dynamic ejecta parameters are: ejecta mass ($M_{\rm ej}=0.01~M_{\sun}$), opacity ($ \kappa=4.0~\rm cm^{2}~\rm g^{-1}$), ejecta velocity ($ \beta=0.11$). The parameters adopted to calculate r-process powered kilonova are $M_{\rm ej}=0.018~M_{\sun}$,$ \kappa=2.5~\rm cm^{2}~\rm g^{-1}$, $ \beta=0.14$.

In figure~\ref{fig:KN}, we can see that the 10 day ($\sim10^{6}$s) excess data can also be well explained by the BP-powered kilonova model. Therefore,  the kilonova emission may provide a further evidence of a newborn BH.

\section{Conclusion and discussion}
The merger of binary neutron stars is an important source of gravitational waves. The product of such a merger may be a stellar mass BH, a rapidly spinning supra-massive magnetar (which may collapse to a black hole after losing centrifugal support), or a stable NS \citep{Gao+2016}.  In this paper, we analyzed multi-band data of GRB~160821B and found that the early X-ray afterglow shows an ``internal plateau'' followed by a sharp decay (Figure~\ref{160821B}), which is often interpreted as the signature of a magnetar engine. This implies that a supra-massive neutron star survive from the merger. A newborn BH is expected from the collapse of the supra-massive NS when it spins down. The fallback accretion of the newborn BH could produce a second X-ray plateau feature if enough energy transfer from the newborn BH to the GRB blast wave. Besides, the injected energy to the ejecta from the BP-driven wind can produce a bright kilonova emission, which provide an additional evidence for the newborn BH. Our model can naturally explain the multi-band afterglow of GRB 160821B as well as kilonova after deducting the optical afterglow component, suggesting that the GRB 160821B-like source may contain a newborn BH.

The fitting of kilonova depends on the afterglow subtraction. In GRB 160821B, the early ($\sim$0.08 day) optical data is dominated by afterglow. As X-ray and optical afterglow is typically in the same spectral regime, the trend of the later afterglow will be determined. We also noticed that \citet{Lamb+2019} has proposed a refreshed shock and two-component kilonova model to explain the optical data. This is essentially an additional energy injection at the late time.

We just employed a simple model to fit the kilonova of GRB160821B. There are several effects should be included. Recently, \citet{Korobkin+2020} showed that ejecta morphologies have significant effects on the light curves of kilonova. The predicted light curves also varied with the viewing angles \citep{Zhu+2020}. Moreover, the neutron-rich outflows from disk could heat up the merger ejecta, and were considered as an additional energy source for kilonova emission \citep{Song+2018}. Therefore, the ejecta mass inferred in our work would involve some uncertainties due to the simple model used.

The GW event with short GRB emissions and followed-up X-ray ``internal plateau'' simultaneously detected would provide a good test to our model. Recently, the Advanced LIGO/Virgo detector observed a compact binary coalescence (GW190425) with the total mass$\sim3.4 M_\sun$ and 90 percent credible intervals for the component masses range from 1.12 to 2.52 $M_\sun$ \citep{Abbott+2020}. \citet{Han+2020} suggested that GW190425 is a candidate of a NS-BH merger event. The inferred BH mass($\sim2.4 M_\sun$) can fill the mass gap ($\sim2-5 M_\sun$) given by X-ray binary observations. It's worth noting that the newborn BH formed as the result of the collapse of a supra-massive magnetar would provide such a low-mass BH.

\acknowledgments

We thank He Gao, A-Ming Chen and Kai Wang for helpful discussion. This work is supported by the National Key R\&D Program of China (Nos. 2020YFC2201400 and 2018YFA0404204), the National Natural Science Foundation of China under grants 11773010, U2038107, 11833003 and U1931203. B.B.Z acknowledges support by the Fundamental Research Funds for the Central Universities (14380035), and the Program for Innovative Talents, Entrepreneur in Jiangsu.

\end{document}